\documentclass[conference]{IEEEtran}
\IEEEoverridecommandlockouts

\usepackage{amsmath,amssymb,amsfonts}
\usepackage{algorithmic}
\usepackage{graphicx}
\usepackage{textcomp}
\usepackage{siunitx}
\usepackage[acronym]{glossaries}
\usepackage{xcolor}
\usepackage{float}
\usepackage[]{todonotes}            % for \todo{}
\presetkeys{todonotes}{inline}{}  % make all \todo inline

\usepackage{caption}

 \usepackage{eso-pic}
 \usepackage{color}
 \definecolor{DisclaimerGray}{gray}{0.92}
 
 \AddToShipoutPicture*{\put(45,760){\colorbox{DisclaimerGray}{\parbox{\textwidth}
 {\footnotesize \copyright IEEE, 2020. This is the author's version of the 
 work. Personal use of this material is permitted. However, permission to 
 reprint/republish this material for advertising or promotional purpose or for 
 creating new collective works for resale or redistribution to servers or lists, 
 or to reuse any copyrighted component of this work in other works must be 
 obtained from the copyright holder. The definite version will be published at  
2020 IEEE International Instrumentation and Measurement Technology Conference (I2MTC)
.}}}}

\begin{document}
\widowpenalty 10000
\clubpenalty 10000

\title{Augmenting an Assisted Living Lab with Non-Intrusive Load Monitoring}

\author{\IEEEauthorblockN{
                             Hafsa Bousbiat}
                            \IEEEauthorblockA{\textit{University of Klagenfurt} \\
                            Klagenfurt, Austria \\
                            hafsa.bousbiat@aau.at}
\and
        \IEEEauthorblockN{ Christoph Klemenjak}
                             \IEEEauthorblockA{\textit{University of Klagenfurt} \\
                            Klagenfurt, Austria \\
                            klemenjak@ieee.org }

\and
        \IEEEauthorblockN{ Gerhard Leitner }
        \IEEEauthorblockA{\textit{University of Klagenfurt} \\
        Klagenfurt, Austria \\  
        Gerhard.Leitner@aau.at}
        \and
        \IEEEauthorblockN{  Wilfried Elmenreich}
                             \IEEEauthorblockA{\textit{University of Klagenfurt} \\
                            Klagenfurt, Austria \\
                            wilfried.elmenreich@aau.at}
}

\maketitle

\maketitle

\begin{abstract}

%The increasing number of electric devices in today's homes is constantly calling for new solutions that allow users to efficiently manage their energy consumption. Real-time feedback would be of a great benefit in this case. \gls{nilm} systems are widely used for this purpose as they offer a good trade-off between the cost and the complexity. 

%%%%%%%%%% the above sentence should be changed
%We further describe a possible implementation of a measurement architecture that supports \gls{nilm} using an open source platform that allows to control and monitor the load of a household. 

The need for reducing our energy consumption footprint and the increasing number of electric devices in today's homes is calling for new solutions that allow users to efficiently manage their energy consumption. Real-time feedback at device level would be of a significant benefit for this application. In addition, the aging population and their wish to be more autonomous have motivated the use of this same real-time data to indirectly monitor the household's occupants for their safety. By breaking down aggregate power consumption into its components, Non-Intrusive Load Monitoring provides information on individual appliances and their current state of operation. Since no additional metering equipment is required, residents are not confronted with intrusion into their familiar environment. Our work aims to depict an architecture supporting non-intrusive measurement with a smart electricity meter and the handling of these data using an open-source platform that allows to visualize and process real-time data about the total energy consumed. As a case study, we describe a series of measurements from common household devices and show how abnormal behavior can be detected.
\end{abstract}


%%%%%%%%%%%%% Acronym Definition %%%%%%%%%%%%%%%%%%%%%%%%

\newacronym{nilm}{NILM}{Non-intrusive Load Monitoring}
\newacronym{ilm}{ILM}{Intrusive Load Monitoring}
\newacronym{aal}{AAL}{Active and Assisted Living}
\newacronym{hmm}{HMM}{Hidden Markov Model}
\newacronym{rnn}{RNN}{Reccurent Neural Networks}
\newacronym{dae}{dAE}{Denoising Auto-Encoder}
\newacronym{lstm}{LSTM}{Long Short Term Memory}
\newacronym{cnn}{CNN}{Convolutional Neural Network}
\newacronym{rcnn}{RCNN}{Recuurent Convolutional Neural Network}
\newacronym{gru}{GRU}{Gated Reccurent Unit}
\newacronym{drn}{DRN}{Deep Residual Network}
\newacronym{han}{HAN}{Home Area Network}
\newacronym{hems}{HEMS}{Home Energy Monitoring systems}
\newacronym{ilm}{ILM}{Intrusive Load Monitoring}
\newacronym{adl}{ADLs}{Activity of Daily Living}
\newacronym{openhab}{openHAB}{open Home Automation Bus}
\newacronym{osgi}{OSGi}{Open Source Gateway initiative}
%%%%%%%%%%%%%%%%%%%%%%%%%%%%%%%%%%%%%%%%%%%%%%%%%%%%%%%%%

%\todo{TODO: find new title; WIL: current title (AaALLwNILM) fine for me; OK, so shall it be this one :) }

\section{Introduction} \label{Introduction}

     Both the excessive use of fossil fuels, as well as the ever-growing demand for energy are contributing to the biggest challenges that
     we are facing today. For example, the worldwide energy consumption grew by 2.3\% in 2018, which is the highest growth rate in this decade (IEA 2018)\footnote{https://www.iea.org/geco/}. With 30 to 40\%, residential homes and buildings account for a significant portion of the overall energy consumption\cite{HEINONEN2014295}. Therefore, we aim at increasing awareness about energy consumption among residents with a special focus on electric energy and smart metering.
    
    The first step towards a more efficient energy consumption behavior would be to identify the greediest appliances in a household. Unfortunately,
    the traditional billing system, where the user is informed about his consumption a few months later, does not allow such a practice. Users generally
    are not able to remember which devices were used in the duration referred in the energy bill and due to having only an overall lump sum it is almost impossible to make connections between the appliances' usage and the resulted power 
    consumption value. Providing  users with real-time information would be, in this case, of a great impact on their behaviors and decisions.
    However, providing only the total consumption does not give much information. Device-level information provides a more clear image
    about the appliances and their consumption. Several works have shown that informing the household's occupants about the energy consumption
    of single appliances can help to lower their total energy consumption by 14\% \cite{ehrhardt2010advanced}. 
    Monacchi et al. \cite{monacchi2017open} proposed an advisor to give personalized feedback to users. As a result, a potential of 34\% of savings has been identified.
     
    In addition to the goal of improving energy efficiency, energy consumption data can also be used to support elderly people to live independently.
    As a matter of fact, today's transition towards a much older population, mainly in the developed countries, is challenging existing models of social support and quality
    of life for older people and their carers.  With the emerging efforts in the field of energy management systems, new opportunities to join these two domains are rising to the surface.
    At first glance, these two domains may seem to be diverging. However recent studies \cite{ruano2019nilm}, have shown that many health features can be inferred from the energy consumption 
    data, examples include sleep disorder, anomalies in activity patterns, etc.
    By drawing the full image of the daily activities, appliance level information can be used by the smart home system in order to detect unusual behavior and trigger medical
    intervention when needed.

     In this paper, we present a straightforward and cost-efficient implementation of an energy measurement system for a Smart
     Home based on an open-source platform for home automation - \gls{openhab}\footnote{https://www.openhab.org/} -  that is intended to be used in future work to 
     detect abnormal patterns in the behavior of the household's occupant. 
     
The remainder of this paper is organized as follows: Section \ref{sec:related} discusses related work on \gls{nilm} and \gls{aal}. Section \ref{Lab} describes the lab environment, which is used as an experiment site. Section \ref{sec:architecture} illustrates the implemented architecture and its components. In Section \ref{sec:studies}, we present two case studies that depict the advantages of combining NILM and AAL. Section \ref{sec:conclusions} concludes the paper and gives an outlook on future work.
    
    \section{Related Work} \label{sec:related}

Improving energy efficiency has received much attention in recent years due to its important economic and ecologic benefits. However, providing a single meter for each appliance (or using exclusively smart appliances~\cite{elmenreich:wises12}) would not be an economic solution because of the cost for procurement, installation, and operation of the metering infrastructure. \gls{nilm} techniques offer a cost-effective solution to monitor devices with a single metering unit. \gls{nilm}, introduced in \cite{hart1992nonintrusive}, estimates the power consumption of individual devices given their aggregate consumption. In this way, the combined consumption must only be monitored at a single, central point in the household, providing various advantages such as reduced cost for metering equipment \cite{klemenjak2020towards}. Certified smart meters or low-cost energy monitors are typically installed to provide information on energy consumption and related quantities of interest \cite{klemenjak2018yomopie}.

Besides the provision of simple feedback, \gls{nilm} is seen as enabling technology for occupancy detection \cite{kleiminger2013occupancy} \cite{chen2013non}. Furthermore, unusual power consumption patterns of appliances can be used to detect faulty appliances or malfunctions. A comprehensive review of \gls{nilm} techniques can be obtained from \cite{faustine2017survey}. 

Besides common application areas such as occupancy detection and feedback on electricity consumption, \gls{nilm} gained momentum in \gls{hems} and \gls{aal} \cite{hernandez2019applications}. \gls{aal} comprises assisted living technologies that rely on ambient intelligence. Among other applications,  \gls{aal} tools are used to cure and improve wellness and health conditions of adults \cite{rashidi2012survey}. Such tools rely on data to draw conclusions, give alerts, or notify ambulance. As described in a recent contribution, NILM can provide key information such as the on/off status of certain household appliances. Based on this information,  \gls{aal} systems can make further important decisions \cite{ruano2019nilm}.

In \cite{alcala2015activity}, the authors present an activity monitoring system for elderly people that leverages information from a \gls{nilm} approach. The presented system aims to classify the observed daily activity in terms of normality. In \cite{alcala2017assessing}, \gls{nilm} information was combined using the Dempster-Shafer theory.

%\todo{TODO: Is there acutally any other lab that integrated NILM into their smart home lab or do we have a unique selling point?}

Not exclusively but also with regard to \gls{aal}, the accuracy of \gls{nilm} algorithms can be improved by incorporating sensor input different than electric power. In \cite{shahriar2013applying}, the authors observed that Machine Learning methods for \gls{nilm} show an increased classification accuracy of up to 38 \% when considering ambient features. Further improvements to performance can be achieved by the use of low-cost sensor networks \cite{le2017improving}. In contrast to installing new sensors, the prototype presented in \cite{berges2010leveraging} makes use of pre-existing environmental sensors of a building. Such environmental sensors provide appliance-related information that results in improved performance of \gls{nilm} algorithms. The authors approve a positive impact of data provided by environmental sensors.

\section{Experiments Site} \label{Lab}

%\todo{TODO: Is there some website or any oder illustration of the lab? I think NILM + this lab is a strong selling point}

   As was mentioned in Section \ref{Introduction}, state of the art technology bears great potential in several areas of application. However, in current solutions, one of the most important involved groups - the end consumers - is not appropriately informed about the possibilities or involved in the development processes. In order to demonstrate the former and better involve users within a \emph{user-centered design approach}, our institution has established a lab facility shaped like an apartment with a living room, a kitchen and an anteroom area\footnote{https://www.aau.at/tewi/infrastruktur/}. The lab allows to run simulations, test smart home components and evaluate prototypical solutions with users. 
   
   The lab is equipped with a combination of smart devices from different manufacturers. Beside others, a Philips hue bulb or a Samsung
    cleaning robot is present. The variety of devices that can be controlled is big. In the kitchen area a stove, a cooking hood, a 
    microwave oven and a refrigerator can be utilized where the first three appliances are all equipped with sensors and actuators.
    Moreover, water consumption can be monitored by a flow meter. In the living room area a smart speaker system, a projector, and 
    a motor-controlled projection wall are present. An integrated sensor is present in the projection wall giving information about
     its state (up / down). In the anteroom, the functions
    and features of smart switches and motion detectors are demonstrated. 
    However, enhancing the enumerated appliances with 
    \emph{smart} functionality is not an easy task, because no system available (speaking of the end consumer market) would be able
    to cover all desired functions and support all appliances. In previous projects \cite{leitner2014disseminating} we have 
    established our own layered architecture based on \gls{osgi}\footnote{The \gls{osgi} specification describes a modular system and a service platform for the Java programming language that implements a complete and dynamic component model} to integrate devices from different manufacturers in one platform.
    Meanwhile, a big Smart Home developer community has established open-source platforms covering the integration of a number
    of smart home systems and components from different manufacturers, providing needed functionality to a great extent and
    offering interfaces and tools to establish customized solutions. The platform in use in our environment is \gls{openhab}, 
    a system based on the eclipse smart home framework. \gls{openhab} provides the optimal software framework for scientific 
    research in the field of Smart Home. As one of the features, \gls{openhab} enables easy storage of device data on standard 
    databases such as \textit{MySQL} for further analyses. Simply defined rules help in simulating automated functions. 
    Both features are not typical for systems available on the end consumer market.  The \gls{openhab} instance is locally installed in the 
    lab on a central unit (All-in-One PC ASUS Eee Top A6420-BF016M) running a Linux platform that can interact with the above-mentioned 
    appliances through the local lab network.      
   
   The main focus of the laboratory in the recent past has been to demonstrate the possibilities of smart home systems in the context of \gls{aal}, for example by showing the possibility to remotely control even conventional devices. This does not only enhance the comfort but also the security, for example by automatically switching off dangerous devices such as the kitchen stove after a certain time. The additional benefit of the basic infrastructure is (as has been mentioned above) the possibility to track and identify \emph{typical} behavior and react when there are significant deviations. However, in state-of-the-art smart home systems, it is still necessary to equip each device in a home with a separate component (sensor or actuator) to be able to track utilization. This requires, in some cases, to intervene in the household's wiring which causes reasonable efforts and is costly. With \gls{nilm} the tracking is possible with comparatively low installation efforts.          
   
\section{System Architecture}

\label{sec:architecture}      
      
\begin{figure*}[tb]
\centering
  \includegraphics[width=0.9\textwidth]{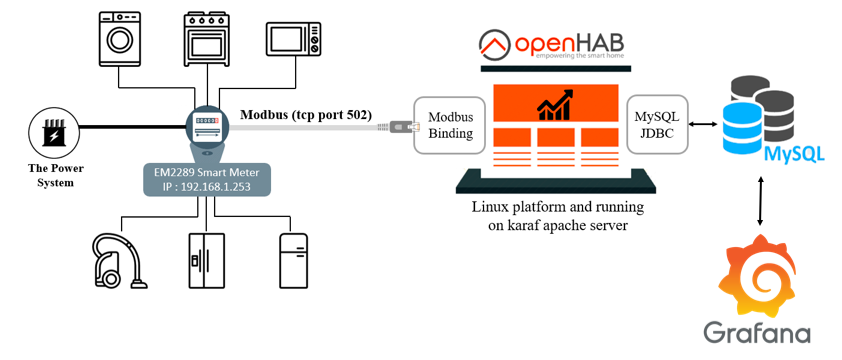}
  \vspace{-1.5em}
  \caption{Overview of the implemented  architecture}
  \label{img:globalArchitect}
\end{figure*}

A recent contribution brought forward 17 suggestions for the collection of energy data. With regard to data collection and data storage, we put a strong emphasis on meeting compliance with those suggestions in the design of our system \cite{klemenjak2019datasets}. The architecture of our system can be divided into two main groups: hardware and software.

The hardware architecture is mainly responsible for data acquisition, whereas the software part handles communication with a smart meter and serves as a data logger. Figure \ref{img:globalArchitect} illustrates our system and interactions between individual components.

\subsection{Hardware Architecture }
    
 %    \begin{figure}[H]
  %      \includegraphics[scale=0.37]{image/HardArchitect.PNG}
   %     \caption{Overview of the hardware architecture.}
    %    \label{img:hardarchitect}
    %\end{figure}

Data collection is a fundamental task of every data acquisition unit and can be decisive for the quality of extracted data. To meet safety guidelines and ensure a predictable measurement error, we integrate a certified smart meter\footnote{Gossen EM2289} in our architecture. Besides compliance with safety and quality standards, this meter can easily be integrated into a \gls{han} through the built-in Ethernet connection. % with a sampling rate of 32 times per period. 
In addition, the Ethernet interface permits remote read-out and data transfer to external processing units. We decided to integrate a smart meter of a European company since it is expected that 72\% of European consumers will have such a smart meter installed in their homes by the end of 2020 \footnote{https://ses.jrc.ec.europa.eu/smart-metering-deployment-european-union}.  
    
%In our simulation, we connected the EM 2289 smart meter to three phases to reach full operation mode. However,  energy consumption of the different appliances was measured only on one phase, using the 9th pin (i.e L3 to N) of the smart meter.
 
%\todo{TODO: describe where the server is located. What does the Linux platform look like? Could we run NILM on it? or even Tensorflow? : done in the expriments sites}

%\todo{TODO: What sensors are in the lab? : done in the expriments sites, Which ones could be use for NILM? }

%\todo{TODO: What quantities can we measure? Active power, current,...? At what rate will data be available? How could a NILM algorithm access the obtained data? Does the architecture allow us to build a smart platform with AI features?}

    \subsection{Software Architecture}
    
%\todo{Reviewer 1: For further detail, authors could provide more discussion about the openHAB tool. Why is it used? why not others?}

\gls{openhab} represents the centerpiece of our software architecture. It can be best described as a home automation platform that allows controlling a variety of devices. The platform offers the scheduling of events and the introduction of a fixed set of rules.
%%%%%%%% Why is it used? why not others? 
%\textcolor{red}{
 It also can operate as a high-security intranet for the sensors and actuators available in the smart home by providing the possibility to process data locally. 
 Besides, unlike other platforms (e.g. ioBroker\footnote{https://www.iobroker.net/},  Home Assistant\footnote{https://www.home-assistant.io/},...
 ), \gls{openhab}  benefits from a robust community support and very clear available documentation.
%}

In our setup, \gls{openhab} acquires power readings from the smart meter via Ethernet. The platform uses a ModBus TCP protocol as networking protocol, which operates in a master/slave mode. In our architecture, the smart meter serves as slave i.e. it waits for requests. The \gls{openhab} unit acts as ModBus master. During the setup, we extended the \gls{openhab} platform with a ModBus binding by specifying:

\begin{itemize}
    \item the IP address of the slave
    \item targeted data (active power, reactive power, ...)
    \item refresh rate (sampling frequency) as \SI{1}{\hertz}
\end{itemize}

%\todo{TODO: is 1 Hertz correct?, what are we reading? active power?}

Though \gls{openhab} integrates a dashboard that updates the latest readings in real time, we integrate a \textit{MySQL} database to serve as data store. In the current version, the database serves as an interface for \gls{nilm} algorithms. Based on readings provided by the MySQL database, \gls{nilm} algorithms can be used to infer present appliances. 
Besides data storage, timely data visualization was identified as a design goal. To illustrate activities in the monitored lab, we utilize a \textit{Grafana} instance. This tool allows communicates with the MySQL database, reads data and provides insights on energy consumption of the lab as well as ambient features.

%\todo{Reviewer 1: More information might be given about the NILM techniques implemented, as well as how the appliances are identified automatically.}

%\textcolor{red}{
It is planned to customize this tool to give detailed insights on activities on the basis of information provided by \gls{nilm} algorithms. 
At the moment, our system utilizes the autonomous load disaggregation approach presented in \cite{egarter2015autonomous}. This approach operates unsupervised and doesn't rely on a priori knowledge about appliances. As this algorithm suffers from moderate labeling accuracy, an important objective of future work is to embed more sophisticated NILM algorithms such as the online particle filter presented in \cite{egarter2014paldi} in our system.
%}

  %  \begin{figure}[]
   % \begin{center}
    %
     %\includegraphics[scale=0.15]{image/Grafana.png}
  %\caption{Grafana's Energy Consumpion Dashboard.}
  %\label{img:Grafana}
   %     
%    \end{center}{}
 %
%\end{figure}   

\begin{figure*}
    \centering
    \includegraphics[width=0.9\textwidth]{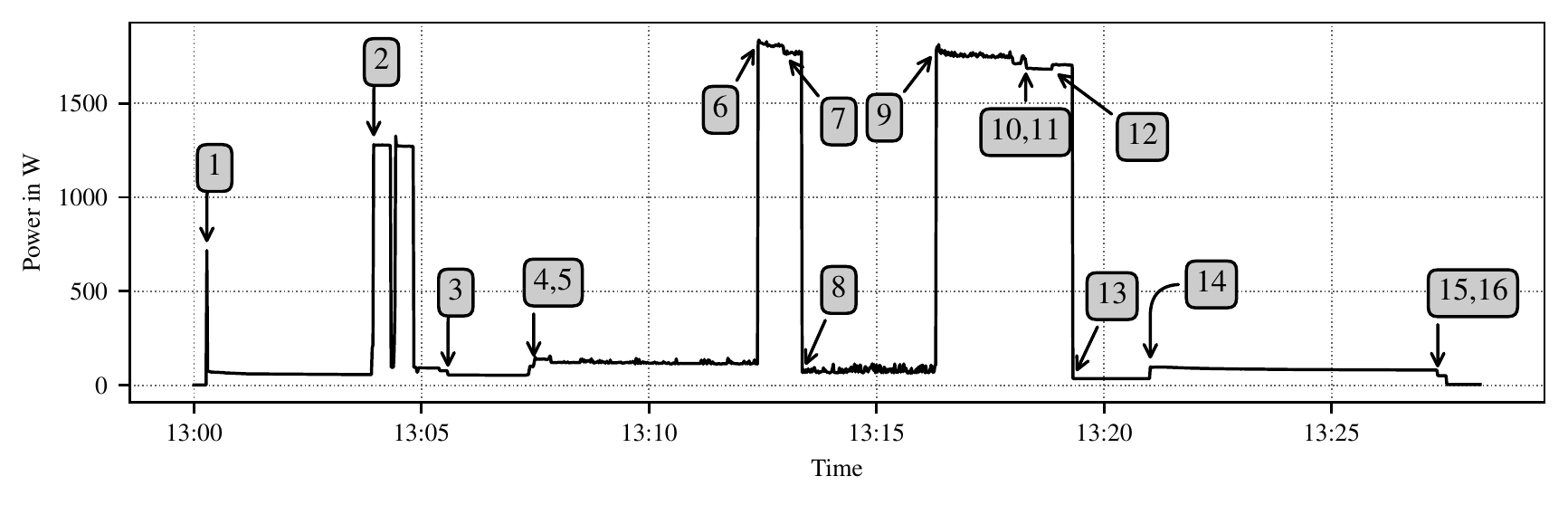}
    \caption{Appliance events in aggregate power consumption}
    \label{fig:exp1}
\end{figure*}

\begin{table}[]
\centering
\begin{tabular}{ccc}
\hline
ID & Appliance & Event Type \\ \hline
1& Fridge &ON \\      
2& Microwave &ON \\
3 & Microwave & OFF \\ 
4 & Monitor & ON \\ 
5 & PC & ON \\ 
6 & Water kettle & ON \\ 
7 & Fridge & OFF \\ 
8 & Water kettle & OFF \\ 
9 & Water kettle & ON \\ 
10 & PC & OFF \\ 
11 & Monitor & OFF \\ 
12 & Ventilator & ON \\ 
13 & Water kettle & OFF \\ 
14 & Fridge & ON \\ 
15 & Ventilator & OFF \\ 
16 & Fridge & OFF \\ \hline
\end{tabular}
\caption{Appliance events in aggregate power readings}
\label{tab:scenario2}
\end{table}

\section{Case Studies}  
\label{sec:studies}

To demonstrate how \gls{nilm} techniques can contribute to scholarship in ambient assistive labs, we present two descriptive case studies: one on automatic labeling of appliance events and the other one on detecting abnormal behavior. 

In the first case study, we simulate a typical household scenario in our lab. This scenario contains appliance usage of six common electrical appliances: fridge, microwave, monitor, a PC, water kettle, and ventilator. Table \ref{tab:scenario2} summarizes appliances in our study, events, and event types.
During a time window of approximately \SI{35}{\minute}, we reenacted behavior within a real household by using appliances to perform specific tasks such as heating water in the kettle, using the PC to edit a document and reheating food in the microwave oven.
Figure \ref{fig:exp1} shows the total power consumption of the lab during our scenario. Table \ref{tab:scenario2} explains the mapping of IDs and appliances. We clearly see how appliance events impact the aggregate power signal. For instance: the turn-on event of the fridge (event 1) is accompanied by a characteristic turn-on spike. ON and OFF events of the water kettle and the microwave result in considerable steps. In contrast, identifying events related to PC, monitor or ventilator represents a bigger challenge because of their low power consumption. As a result of knowledge learned during training, \gls{nilm} algorithms can provide further insights and solve issues such as detecting appliance with problematically low power consumption. Furthermore, \gls{nilm} techniques can help in understanding the current type of activity in households or reveal the routines of residents. Having information about the routines of the resident is  of great importance in home automation, for example, to switch off appliances that were unintentionally left on and have many use-cases in \gls{aal}.

\begin{figure}
    \centering
    \includegraphics[width=0.9\columnwidth]{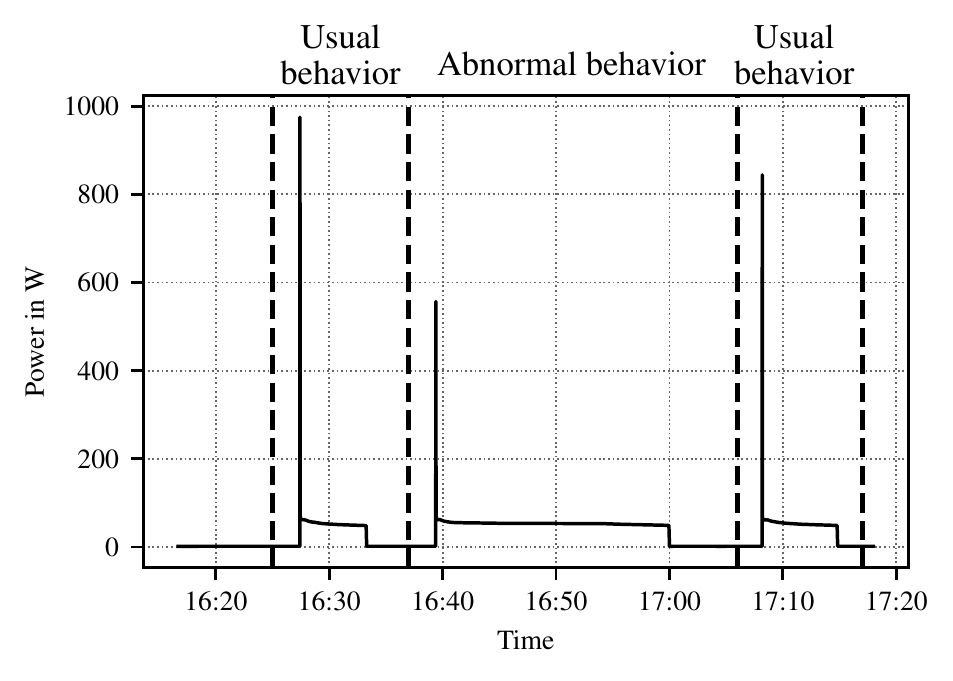}
    \caption{An example of abnormal behavior}
    \label{fig:exp2}
\end{figure}

In the scenario, we want to emphasize a major advantage of \gls{nilm} for \gls{aal}: extraction of individual load curves. The majority of \gls{nilm} algorithms reconstructs the power consumption curve of appliances of interest by observing the aggregate consumption. In this study, we assume that a \gls{nilm} algorithm provides the load curve illustrated in Figure \ref{fig:exp2}. In this figure, we observe three duty cycles of the fridge. Usually, fridges show a periodical behavior that is the result of a cooling process that is initiated when the temperature drops below a certain level and turns off when the target temperature is reached. At the beginning of the measurements, the power consumption shows a periodic behavior with a predictable power consumption pattern. Then, a different cycle can be observed between 16:35 and 17:05. As Figure \ref{fig:exp2} shows, the pattern is elongated to an uncommon extent. As it turns out, this anomalous pattern was caused by having the door of the fridge left open. Consequently, the fridge altered its common behavior and produces an anomaly. Such anomalies can be detected by \gls{aal} tools, which could suggest countermeasures to the causes of  anomalies or, in case of emergency, notify rescue services. This simple scenario shows how \gls{aal} and related applications can benefit from \gls{nilm} algorithms. 

The two case studies presented in this section  clearly demonstrate how \gls{nilm} techniques can assist for deeper insights into activities in homes to assist AAL tools and eventually, improve quality of services. The main limitation of this approach is its inability to detect anomalies in the behavior that is not related to electrical devices. One possibility to overcome this issue would involve installation of additional sensors, which is the opposite of what \gls{nilm} tries to achieve. We conclude a trade-off between sensor installation and technical requirements has to be found.

\section{Conclusion} \label{sec:conclusions}
This work presented  a proof-of-concepts realization of an in-door smart energy measurement system. The general architecture features a single smart meter which is connected to an \gls{openhab} system running on a Linux platform on a local server in the lab. The system is able to process 
%and disaggregate
measurement data and to visualize information for users. Through the integration of load disaggregation techniques, the presented system is expected to provide deeper insights into activities in the lab and improved diagnostics. In the first set of experiments, energy data was measured to depict appliance consumption patterns and to identify abnormal behavior of appliances like a fridge with an open door. The presenting approach is a promising approach for future \gls{aal} systems that support users in increasing energy awareness, reducing energy consumption, or supporting elderly people in their active living at their own home.
Future work will address the testing of different load disaggregation algorithms to automatically identify the active devices using only the total power. Further studies could also investigate the combination and visualization of the data provided by our measurement architecture and other sensors present in the smart Home to better detect abnormal behavior.

%\todo{TODO: What do you think will be done next?}

\section{Acknowledgements}

This work was supported by the Doctoral College DECIDE and by Lakeside Labs via the Smart Microgrid Lab.

%\printbibliography{}



\begin{thebibliography}{10}
\providecommand{\url}[1]{#1}
\csname url@samestyle\endcsname
\providecommand{\newblock}{\relax}
\providecommand{\bibinfo}[2]{#2}
\providecommand{\BIBentrySTDinterwordspacing}{\spaceskip=0pt\relax}
\providecommand{\BIBentryALTinterwordstretchfactor}{4}
\providecommand{\BIBentryALTinterwordspacing}{\spaceskip=\fontdimen2\font plus
\BIBentryALTinterwordstretchfactor\fontdimen3\font minus
  \fontdimen4\font\relax}
\providecommand{\BIBforeignlanguage}[2]{{%
\expandafter\ifx\csname l@#1\endcsname\relax
\typeout{** WARNING: IEEEtran.bst: No hyphenation pattern has been}%
\typeout{** loaded for the language `#1'. Using the pattern for}%
\typeout{** the default language instead.}%
\else
\language=\csname l@#1\endcsname
\fi
#2}}
\providecommand{\BIBdecl}{\relax}
\BIBdecl

\bibitem{HEINONEN2014295}
J.~Heinonen and S.~Junnila, ``Residential energy consumption patterns and the
  overall housing energy requirements of urban and rural households in
  finland,'' \emph{Energy and Buildings}, vol.~76, pp. 295 -- 303, 2014.

\bibitem{ehrhardt2010advanced}
K.~Ehrhardt-Martinez, K.~A. Donnelly, S.~Laitner \emph{et~al.}, ``Advanced
  metering initiatives and residential feedback programs: a meta-review for
  household electricity-saving opportunities.''\hskip 1em plus 0.5em minus
  0.4em\relax American Council for an Energy-Efficient Economy Washington, DC,
  2010.

\bibitem{monacchi2017open}
A.~Monacchi, F.~Versolatto, M.~Herold, D.~Egarter, A.~M. Tonello, and
  W.~Elmenreich, ``An open solution to provide personalized feedback for
  building energy management,'' \emph{Journal of Ambient Intelligence and Smart
  Environments}, vol.~9, no.~2, pp. 147--162, 2017.

\bibitem{ruano2019nilm}
A.~Ruano, A.~Hernandez, J.~Ure{\~n}a, M.~Ruano, and J.~Garcia, ``{NILM}
  techniques for intelligent home energy management and ambient assisted
  living: A review,'' \emph{Energies}, vol.~12, no.~11, p. 2203, 2019.

\bibitem{elmenreich:wises12}
W.~Elmenreich and D.~Egarter, ``Design guidelines for smart appliances,'' in
  \emph{Proceedings of the 10th International Workshop on Intelligent Solutions
  in Embedded Systems (WISES'12)}, Klagenfurt, Austria, Jul. 2012.

\bibitem{hart1992nonintrusive}
G.~W. Hart, ``Nonintrusive appliance load monitoring,'' \emph{Proceedings of
  the IEEE}, vol.~80, no.~12, pp. 1870--1891, 1992.

\bibitem{klemenjak2020towards}
C.~Klemenjak, S.~Makonin, and W.~Elmenreich, ``Towards comparability in
  non-intrusive load monitoring: On data and performance evaluation,'' in
  \emph{2020 IEEE Power \& Energy Society Innovative Smart Grid Technologies
  Conference (ISGT)}, 2020.

\bibitem{klemenjak2018yomopie}
C.~Klemenjak, S.~Jost, and W.~Elmenreich, ``{YoMoPie: A} user-oriented energy
  monitor to enhance energy efficiency in households,'' in \emph{2018 IEEE
  Conference on Technologies for Sustainability (SusTech)}.\hskip 1em plus
  0.5em minus 0.4em\relax IEEE, 2018, pp. 1--7.

\bibitem{kleiminger2013occupancy}
W.~Kleiminger, C.~Beckel, T.~Staake, and S.~Santini, ``Occupancy detection from
  electricity consumption data,'' in \emph{Proceedings of the 5th ACM Workshop
  on Embedded Systems For Energy-Efficient Buildings}.\hskip 1em plus 0.5em
  minus 0.4em\relax ACM, 2013, pp. 1--8.

\bibitem{chen2013non}
D.~Chen, S.~Barker, A.~Subbaswamy, D.~Irwin, and P.~Shenoy, ``Non-intrusive
  occupancy monitoring using smart meters,'' in \emph{Proceedings of the 5th
  ACM Workshop on Embedded Systems For Energy-Efficient Buildings}.\hskip 1em
  plus 0.5em minus 0.4em\relax ACM, 2013, pp. 1--8.

\bibitem{faustine2017survey}
A.~Faustine, N.~H. Mvungi, S.~Kaijage, and K.~Michael, ``A survey on
  non-intrusive load monitoring methodies and techniques for energy
  disaggregation problem,'' \emph{arXiv preprint arXiv:1703.00785}, 2017.

\bibitem{hernandez2019applications}
{\'A}.~Hern{\'a}ndez, A.~Ruano, J.~Ure{\~n}a, M.~Ruano, and J.~Garcia,
  ``Applications of {NILM} techniques to energy management and assisted
  living,'' \emph{IFAC-PapersOnLine}, vol.~52, no.~11, pp. 164--171, 2019.

\bibitem{rashidi2012survey}
P.~Rashidi and A.~Mihailidis, ``A survey on ambient-assisted living tools for
  older adults,'' \emph{IEEE Journal of Biomedical and Health Informatics},
  vol.~17, no.~3, pp. 579--590, 2012.

\bibitem{alcala2015activity}
J.~Alcal{\'a}, J.~Ure{\~n}a, and {\'A}.~Hern{\'a}ndez, ``Activity supervision
  tool using non-intrusive load monitoring systems,'' in \emph{2015 IEEE 20th
  Conference on Emerging Technologies \& Factory Automation (ETFA)}.\hskip 1em
  plus 0.5em minus 0.4em\relax IEEE, 2015, pp. 1--4.

\bibitem{alcala2017assessing}
J.~Alcal{\'a}, J.~Ure{\~n}a, {\'A}.~Hern{\'a}ndez, and D.~Gualda, ``Assessing
  human activity in elderly people using non-intrusive load monitoring,''
  \emph{Sensors}, vol.~17, no.~2, p. 351, 2017.

\bibitem{shahriar2013applying}
M.~S. Shahriar, A.~Rahman, and D.~Smith, ``Applying context in appliance load
  identification,'' in \emph{2013 Ninth International Conference on Natural
  Computation (ICNC)}.\hskip 1em plus 0.5em minus 0.4em\relax IEEE, 2013, pp.
  900--905.

\bibitem{le2017improving}
X.-C. Le, ``Improving performance of non-intrusive load monitoring with
  low-cost sensor networks,'' Ph.D. dissertation, {\'E}cole Nationale
  Sup{\`e}rieure des Sciences Appliqu{\`e}es et de Technologie, 2017.

\bibitem{berges2010leveraging}
M.~Berg{\'e}s, L.~Soibelman, and H.~S. Matthews, ``Leveraging data from
  environmental sensors to enhance electrical load disaggregation algorithms,''
  in \emph{Proceedings of the 13th International Conference on Computing in
  Civil and Building Engineering, Nottingham, UK}, vol.~30, 2010.

\bibitem{leitner2014disseminating}
G.~Leitner, A.~Felfernig, A.~Fercher, and M.~Hitz, ``Disseminating ambient
  assisted living in rural areas,'' \emph{Sensors}, vol.~14, no.~8, pp.
  13\,496--13\,531, 2014.

\bibitem{klemenjak2019datasets}
C.~Klemenjak, A.~Reinhardt, L.~Pereira, M.~Berges, S.~Makonin, and
  W.~Elmenreich, ``Electricity consumption data sets: Pitfalls and
  opportunities,'' in \emph{The 6th ACM International Conference on Systems for
  Energy-Efficient Buildings, Cities, and Transportation (BuildSys '19)}.\hskip
  1em plus 0.5em minus 0.4em\relax ACM, 2019.

\bibitem{egarter2015autonomous}
D.~Egarter and W.~Elmenreich, ``Autonomous load disaggregation approach based
  on active power measurements,'' in \emph{2015 IEEE International Conference
  on Pervasive Computing and Communication Workshops (PerCom Workshops)}.\hskip
  1em plus 0.5em minus 0.4em\relax IEEE, 2015, pp. 293--298.

\bibitem{egarter2014paldi}
D.~Egarter, V.~P. Bhuvana, and W.~Elmenreich, ``{PALDi}: Online load
  disaggregation via particle filtering,'' \emph{IEEE Transactions on
  Instrumentation and Measurement}, vol.~64, no.~2, pp. 467--477, 2014.

\end{thebibliography}
\end{document}